\documentclass{article}
\usepackage[preprint]{spconf}
\usepackage{amsmath,graphicx}
\usepackage{amssymb,amsfonts,amsthm, booktabs}
\usepackage[colorlinks = true,
            linkcolor = blue,
            urlcolor  = blue,
            citecolor = BlueViolet,
            anchorcolor = blue]{hyperref}

\newcommand{\MYhref}[3][BlueViolet]{\href{#2}{\color{#1}{#3}}}
\usepackage{multirow}
\usepackage[dvipsnames]{xcolor}

\newcommand\todo[1]{} %
\title{Unsupervised Lead Sheet Generation via Semantic Compression}
\name{Zachary Novack \qquad Nikita Srivatsan \qquad Taylor Berg-Kirkpatrick \qquad Julian McAuley}
\address{University of California San Diego}
\copyrightnotice{\footnotesize \begin{tabular}[t]{@{}l@{}} \copyright IEEE 2023. Personal use of this material is permitted. Permission from IEEE must be obtained for all other uses, in any current or future media, including \\reprinting/republishing this material for advertising or promotional purposes, creating new collective works, for resale or redistribution to servers or lists, or\\ reuse of any copyrighted component of this work in other works.\end{tabular}}
\begin{document}
\maketitle
\begin{abstract}
Lead sheets have become commonplace in generative music research, being used as an initial compressed 
representation for downstream tasks like multitrack music generation and automatic arrangement. 
Despite this,
researchers have often fallen back on
deterministic reduction methods (such as the skyline algorithm)
to generate
lead sheets when seeking paired lead sheets and full scores,
 with little attention being paid toward the quality of the lead sheets themselves and how they accurately reflect their orchestrated counterparts. 
To address these issues, we propose 
the problem of
\emph{conditional lead sheet generation} (i.e.~generating a lead sheet \emph{given} its full score version), and show that this task can be formulated as an unsupervised music compression task, where the lead sheet represents a compressed latent version of the score.
We introduce a novel model, called Lead-AE, that models the lead sheets as a discrete subselection of the original sequence, using a differentiable top-k operator 
to allow for 
controllable
local sparsity constraints. Across both automatic proxy tasks 
and direct human evaluations, we find that our method improves upon
the established
deterministic
baseline
and produces 
coherent 
reductions of large multitrack scores.
\end{abstract}
\begin{keywords}
symbolic music generation, lead sheet generation, transformers, latent variable models,  music reduction
\end{keywords}
\section{Introduction}
\label{sec:intro}

\begin{figure*}[th]
    \centering
    \includegraphics[trim=20 200 20 200, width=0.9\textwidth]{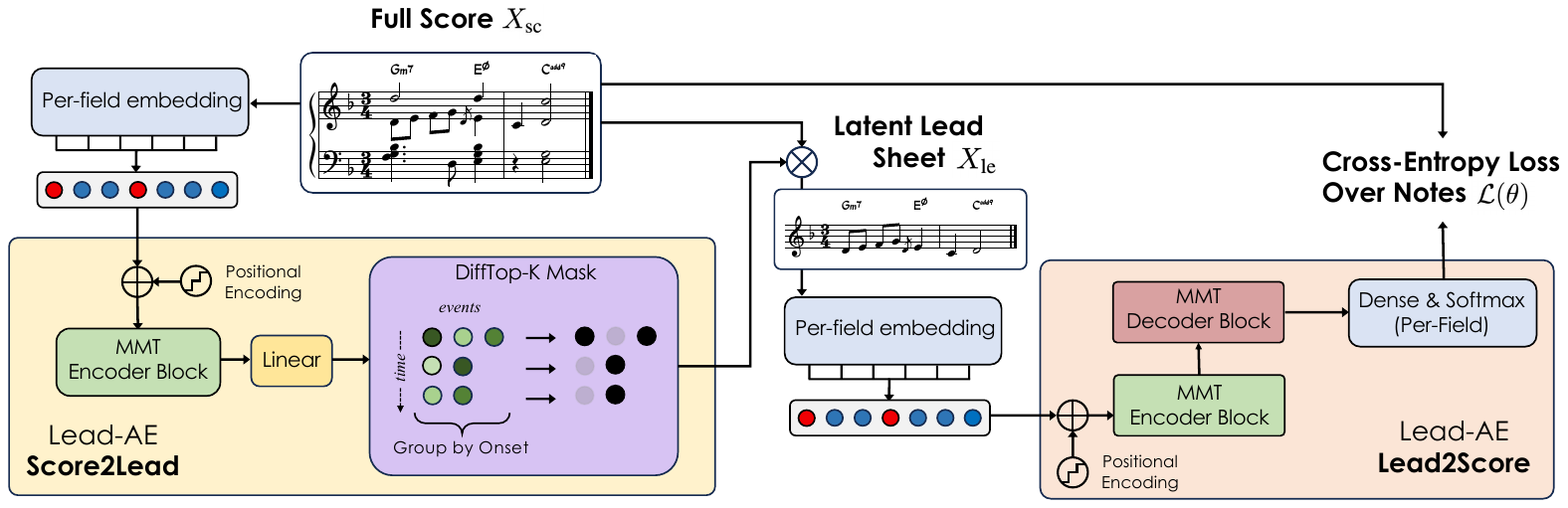}
    \caption{Overview of Lead-AE. Score2Lead uses a transformer encoder and a differentiable top-$k$ mask to select specific notes and chords from the sequence. Lead2Score uses an encoder-decoder transformer to recreate the score given the lead sheet.}
    \label{fig:model}
\end{figure*}

With the goal of communicating high-level musical information quickly and efficiently, \textbf{lead sheets}, or compressed scores that contain
the musically important notes of a larger arrangement
and high-level chord symbols (see Fig.~\ref{fig:model}), are one of the most common musical representations used by modern musicians.
Recently, 
many
in the larger AI Music community
have looked to using lead sheets
for their efficient representation,
whether for generating lead sheets from scratch \cite{de2020rhythm, makris2021generating, wu2020jazz} or using lead sheets as input to larger music generation and arrangement systems \cite{zhao2021accomontage, yi2022accomontage2, wu2023compose, dai2021controllable}. Despite this
growing interest, 
there is a 
data sparsity issue, with 
few publicly available datasets containing high quality lead sheets, and fewer still having paired lead sheet--full arrangement data \cite{wang2020pop909}. This has led many researchers to fall back to deterministic reduction methods like the skyline algorithm when needing paired score--lead sheet data \cite{wu2020jazz, wu2023compose, thickstun2023anticipatory}, and to use these heuristic algorithms for constructing datasets themselves \cite{wang2020pop909}. Furthermore, this trend has exacerbated an over-reliance in generative music on \emph{Pop} music generation and arrangement, 
which contains simpler musical structures that 
algorithms like skyline have shown strong results on
\cite{wu2020jazz,chou2021midibert, hsiao2021learning}, as opposed to more complex music like classical or jazz.
Considering this lack of attention to the quality of the ground truth lead sheets themselves,
this 
raises
the question:
how can we model compressed music (i.e.~lead sheets) that captures the important information from full scores across diverse genres?

In this work we thus propose the 
problem of \emph{conditional lead sheet generation}, where the goal is to generate the  lead sheet given the corresponding full score. Given 
the lack of
paired lead sheets and full scores, 
we design our approach around the intuition that lead sheets carry a sparse selection of notes with the highest information content; thus, we can aim to select a subset of notes and chords from the original score that maximize reconstruction 
of the original score, making the process entirely unsupervised.
Inspired by work in the text domain on discrete latent variable modeling \cite{jhamtani2020narrative, he2020probabilistic, mireshghallah2021style}, we propose \textbf{Lead-AE}, a novel Auto-Encoder--like setup 
wherein the encoder 
leverages
a transformer encoder
with a differentiable top-$k$ operator to target specific notes of the input piece, which are then passed to an
encoder-decoder transformer
to regenerate the initial sequence. On automatic proxy evaluations through reconstruction and melody recovery, as well as both audio and sheet music--based human evaluations, Lead-AE consistent outperforms the strong deterministic baseline.
Our proposed model opens up new avenues for work in conditional lead sheet generation for multitrack music, and can potentially improve the capabilities of existing generation and accompaniment models. Our source code is available at 
\MYhref{https://github.com/ZacharyNovack/Lead-AE}{github.com/zacharynovack/lead-ae}.

\section{Related Work}

 A number of previous works have focused on \emph{unconditional} lead sheet generation (i.e.~from scratch), using both traditional RNN architectures \cite{de2020rhythm, makris2021generating} and modern transformers \cite{makris2021generating, wu2020jazz, zou2022melons, dai2021controllable}. Here, the authors generally rely on either the Wikifonia dataset or the WJazzD \cite{pfleiderer2017inside} dataset, which both only contain lead sheets with no paired scores. Recently, there has been a 
 number of works
 using lead sheets as the first step in a larger music generation system, in both long-form structured music generation \cite{wu2023compose} and accompaniment \cite{zhao2021accomontage, yi2022accomontage2}. In all such works, the lead sheets only serve as an initial ``planning" step for unconditional generation, and are not concerned with how well the lead sheet directly summarizes the full score output.

 Using discrete latent spaces where the input and latent representations share the same structure (e.g.~text) has been well-studied in the NLP community, being used to improve translation tasks \cite{he2020probabilistic}, story generation \cite{jhamtani2020narrative}, and text debiasing \cite{mireshghallah2021style}. To our knowledge, 
 ours
 is the first work to extend this framework to the symbolic music domain, directly modeling our latent space as a symbolic music sequence (rather than an abstract continuous or vector-quantized space).

\section{Proposed Method}

\subsection{Data Representation}

We follow the data representation put forth
by the state-of-the-art Multitrack Music Transformer (MMT) \cite{dong2023multitrack} for efficient modeling of dense multitrack music. Each 
note is represented as a multidimensional input with type, beat, position, pitch, duration and instrument fields,
which reduces the number of tokens needed to represent a given polyphonic sequence (as opposed to tokenizations like REMI+ \cite{von2022figaro}), and 
maintains a 1-to-1 mapping from each musical note to a given token (see \cite{dong2023multitrack} for more detailed information).

In order to model the symbolic chord labels present in lead sheets, 
we use an off-the-shelf chord extractor
based on state-of-the-art work in chord recognition \cite{jiang2019large, mcfee2017structured}. We assume an overly dense set of chord labels, with chord tokens extracted for every beat,
and thus part of the learning problem also includes selecting which chords to keep. 
Every chord 
$\mathbf{c}_i$ is represented as
$\mathbf{c}_i = (c_i^{\text{type}}, c_i^{\text{beat}}, c_i^{\text{position}}, c_i^{\text{root}}, c_i^{\text{quality}}),
$ where $c_i^{\text{root}}$ denotes the pitch class
and $c_i^{\text{quality}}$ denotes the chord type (see Fig.~\ref{fig:model} for an example of chord symbols).

\subsection{Learning Objective}
Conditional Lead Sheet Generation is at its core a sequence translation problem, with the goal of converting a given multitrack music sequence $X_{\text{sc}}$ to a reduced lead sheet sequence $X_{\text{le}}$, where $X_{\text{sc}}$ is the set of original notes $\mathbf{x}_i$ and densely notated chords $\mathbf{c}_i$ and $X_{\text{le}} \subseteq X_{\text{sc}}$. As lead sheets reduce the total number of events from the full score yet maintain a consistent \emph{local} coherence in terms of information density, we require that $X_{\text{le}}$ resides in the constraint set $\mathcal{C}_k(X_{\text{sc}})$:
\begin{equation}\label{eq:k}
    \mathcal{C}_k(X_{\text{sc}}) = \{X_{\text{le}} \subseteq X_{\text{sc}}: |X_{\text{le}}[i \in o]| = k, \forall o \in O\},
\end{equation}
i.e.~that for every unique onset $o \in O$ in the score, $X_{\text{le}}$ can only include $k$ events that share each onset.

To our knowledge there are no large scale datasets of paired lead sheets and full scores available aside from the POP909 \cite{wang2020pop909} dataset, which only contains around 1K songs, is 
limited
to  
pop music, and only has automatically-labeled chords. In order to 
circumvent this,
we note that lead sheets 
can be viewed as a form of ``musical compression": while the form of the data is the same, lead sheets are \emph{semantically} compressed, containing 
only the
most important parts of the original score for easy communication.
Thus, we can formulate conditional lead sheet generation as an \emph{unsupervised} compression task, with the aim of learning a Score2Lead encoder 
to squeeze $X_{\text{sc}}$ into a latent reduction $X_{\text{le}}$ such that a learned Lead2Score decoder 
can reconstruct $X_{\text{sc}}$ as accurately as possible, with the following objective:
\begin{equation}\label{eq:opt}
    \mathcal{L}(\theta) = - \mathbb{E}_{P_{\text{enc}}(X_{\text{le}} \mid X_{\text{sc}})}[\log P_{\text{dec}}(X_{\text{sc}} \mid X_{\text{le}})],
\end{equation}
which is equivalent to the reconstruction term from the standard VAE objective. Note that this expectation is not tractable, as it requires summing over the combinatorial space of lead sheets $X_{\text{le}}$ from $P_{\text{enc}}$ that are in $\mathcal{C}_k$; however, we show in the next section that we can approximate this expectation using a differentiable top-$k$ operator.

\subsection{Lead-AE Formulation}

\textbf{Score2Lead Architecture.}\quad 
Lead sheet generation requires a model to learn which of the $N$ notes 
and $C$ extracted chords
should be included in the latent lead sheet to maximize score reconstruction.
We use an encoder Multitrack Music Transformer (MMT) \cite{dong2023multitrack}
module, where the model attends to the \emph{entire} score input, by taking in the multidimensional sequence as the sum over each field's learned embeddings, with a learnable absolute positional embedding \cite{vaswani2017attention}. 
The
latent outputs $\mathbf{h} \in \mathbb{R}^{(N+C) \times d}$ (where $d$ is the latent dimension) are then passed through a linear layer, that squeezes each latent representation 
into a single score per token $s_i \in \mathbb{R}$.\\[0.3em]
\noindent\textbf{Objective Approximation.} Like in VAEs, the expectation in Eq.~\ref{eq:opt} is intractable. In order to handle this as well as the local sparsity constraints from Eq.~\ref{eq:k},
we approximate
the top-$k$ distribution in each onset using a differentiable top-$k$ operator.
Specifically, after grouping the sequence of $s_i$ by shared note onsets, 
we can apply
the top-$k$ operator 
from
\cite{xie2019reparameterizable}, 
amounting to iterative applications of the Gumbel-Softmax trick 
\cite{jang2016categorical}
along each onset to produce approximate top-$k$ probabilities. We then use the straight-through estimator to convert the soft probabilities to a hard mask, and multiply this with our initial input sequence to generate our latent lead sheet $X_{\text{le}}$, 
which is gauranteed to obey Eq.~\ref{eq:k}
(see the left of Fig.~\ref{fig:model} for an overview of the system). 
We enforce the entire lead sheet to have the same unified instrument 
to further compress the musical representation. At inference time, we deterministically select the top-$k$ notes per onset.\\[0.3em]
\noindent\textbf{Lead2Score Architecture.}\quad The L2S module follows a standard encoder-decoder transformer \cite{vaswani2017attention} using the MMT framework \cite{dong2023multitrack}, where $X_{\text{sc}}$ 
is autoregressively generated given the latent lead sheet
$X_{\text{le}}$. 
Thus, the entire Lead-AE system is trained to minimize $\mathcal{L}(\theta)$ across each field.
\\[0.3em]
\noindent\textbf{Module Pretraining.}\quad
Training the entire Lead-AE system end to end from scratch 
is difficult,
as the approximate gradients of the discrete latent lead sheet can throw off 
early training.
Thus, we 
warm start 
Lead-AE with
the skyline algorithm (which selects the highest pitch note at each onset), 
using the skyline reduction as the input 
for L2S 
and as a supervised target mask for S2L.

\section{Results}

\subsection{Experimental Setup}

Though an optimal assessment of our model would involve evaluating Lead-AE by how well the latent lead sheet 
matches human--produced lead sheets,
no datasets exist to our knowledge that have paired scores and human--annotated lead sheets \emph{including} chords. We thus run our automatic evaluation over two proxy tasks: the overall reconstruction accuracy of the full score using the Symbolic Orchestral Database (SOD) \cite{crestel2018database}, and how well 
Lead-AE
can recover human--annotated \emph{melodies} from POP909 \cite{wang2020pop909}. Beyond this, we turn to two parallel human evaluations, which are able to directly assess the quality of
Lead-AE's outputs.
SOD contains 357 hours of 
multitrack orchestral music, while POP909 contains 60 hours of Chinese pop songs.
We adopt an 0.8/0.1/0.1 train-val-test split for both datasets. Each MMT block 
in Lead-AE
has 4 attention layers with a hidden dimension of 512 and 8 attention heads. We use a maximum sequence length of 1024 and a maximum beat of 256. For all experiments, 
we randomly augment the data
by
shifting the starting beat 
and pitch shifting the sequence within a $\pm 6$ semitone range. All models are trained for a maximum of 1000 epochs, or if the validation performance stagnates for 20 epochs.
For inference, we use top-$k$ sampling and enforce a monotonicity constraint 
for the type and beat fields to ensure coherent outputs.

\subsection{Automatic Evaluation}

\begin{table*}[]
    \centering
    \begin{tabular}{lcc|cccc}
    \toprule
    Model & Note Density& Chord Density& (R) MuTE& (R) PC-MuTE & (R) Jac.  & (R) PC-Jac. \\
    \midrule
    Skyline & 37\% & 100\% & 59.95 & 75.76 & 42.75 & 63.92 \\
    LAE ($k=1$) & 37\% & 100\% & \underline{62.14} &\underline{78.21} & \underline{45.21} & \underline{67.00} \\
    LAE ($k=0.1$) & 39\% & 95\% & \textbf{64.04} & \textbf{79.48} & \textbf{46.59} & \textbf{67.92} \\ 
    \bottomrule
    \end{tabular}
    \caption{Average quantitative results on SOD test set across 3 random seeds, showing the global note and chord densities of the latent lead sheets, and the MuTE and Jaccard results between the original and reconstructed scores.}
    \label{tab:quant}
\end{table*}

We first evaluate Lead-AE's performance on reconstructing the original score from the learned lead sheets 
(as a proxy for evaluating the lead sheets directly).
We use the deterministic skyline algorithm with all chords included as a baseline (i.e.~just training an MMT encoder-decoder model with the fixed skyline lead sheets as input), as it has 
shown
competitive results 
in melody extraction 
\cite{wu2020jazz, chou2021midibert}. For Lead-AE (LAE), we experiment with two varieties: one that maintains the same local sparsity as the skyline algorithm at $k=1$ event per onset (or 2 if there is a chord present), and a more flexible model that uses a \emph{fractional} $k$ at each onset, taking $\lceil10\%\rceil$ of the notes at a given onset to allow the model to adapt to the local density of the score. 
Given previous work on music translation \cite{gover2022music}, we report the reconstruction MuTE metric,
which measures the F1 score per time-step,
as well as the global Jaccard (Jac.) similarity, between the original and reconstructed (R) sequences. We 
additionally report both metrics  
over the output \emph{pitch-classes} (PC), which compresses the sequence to a single octave.
In Table \ref{tab:quant}, Lead-AE outperforms the skyline model with the same sparsity constraint in all metrics, and notably the variable sparsity model (which includes slightly higher note density and lower chord density) dominates both skyline baseline and the fixed sparsity Lead-AE.
Thus, we use the variable sparsity model for all further evaluations.

We also evaluate how well Lead-AE accurately recovers the ground truth melodies in POP909 without 
seeing them during training. 
Since POP909 has long phrases of each song with no melody, we 
focus specifically on the
melody--present sections, 
as both Lead-AE and skyline are designed to pick \emph{which} notes should be included,
not whether 
notes should be included at all.
We finetune both Lead-AE ($k=0.1$) and the skyline baseline on POP909, and report the MuTE scores between the generated \emph{lead sheet} (excluding chords) and the ground truth melodies.
In Table \ref{tab:pop909}, Lead-AE performs as good as skyline (which has strong performance, echoing \cite{wu2020jazz, chou2021midibert}) in recovering the ground truth melodies.
\begin{table}[]
    \centering
    \begin{tabular}{l|cc}
    \toprule
    Model & (LS) MuTE &  (LS) PC-MuTE \\
    \midrule
    Skyline & 77.93 & 81.86\\
    LAE ($k = 0.1$) & \textbf{77.94} & \textbf{82.04}\\
    \bottomrule
    \end{tabular}
    \caption{MuTE scores on POP909 between latent lead sheets and the  ground-truth melodies.}
    \label{tab:pop909}
    \vspace{-0.15cm}
\end{table}
\subsection{Human Listening Study}
While the automatic metrics presented are informative, it is worth noting that they are at best a proxy for human judgement.
Therefore, in order to assess the quality of the generated lead sheets \emph{directly}, 
we conducted 
two rounds of 
human evaluation where annotators were asked to rank the outputs of our system against the skyline baseline based on an audio rendering and the sheet music respectively.
For the listening study, 
we recruited 12 participants at varying levels of musical maturity (from novice to professional musicians).
We randomly selected 11 short excerpts, all from the validation set of SOD. For each excerpt, 
users were asked to listen to an audio rendering of the 
full score, and then were presented both the audio corresponding to the skyline lead sheet and our Lead-AE--generated lead sheet.
We rendered all chord symbols
in root position starting from C3, and all scores were recorded using 
the same
piano patch no matter the instrument, in order to remove timbre differences from the evaluation.
The systems were not labeled and their order was randomized. 

Participants were asked to rank the lead sheets along two parallel criteria: \textbf{(1)} Which lead sheet more \emph{accurately} captured important musical information from the original score? \textbf{(2)} Which lead sheet sounded more musically \emph{fluent} and coherent, irrespective of 
the original?
They were also allowed to answer that the two were too similar to 
distinguish, 
although this was discouraged.
As shown in the left half of Table~\ref{tab:listening},
listeners strongly preferred Lead-AE's reductions over the skyline, and only rarely felt that they were too similar to distinguish.
The proportions were similar for both accuracy and fluency.
Per-song an average of 62.1\% of annotators agreed with the majority answer for accuracy and 60.6\% for fluency, indicating a reasonable degree of inter-annotator agreement.
\begin{table}[]
    \centering
    \begin{tabular}{l|cc|cc}
    \toprule
    \multirow{2}{*}{Model} & \multicolumn{2}{c|}{Listening} & \multicolumn{2}{c}{Reading}\\
    & Acc. & Flu. & Acc. & Flu. \\
    \midrule
    Skyline & 26.5 & 28.0 & 6.7 & 20.0\\
    LAE ($k = 0.1$) & \textbf{57.6} & \textbf{56.8} & \textbf{80.0} & \textbf{56.7}\\
    Too similar & 15.9 & 15.2 & 13.3 & 23.3 \\
    \bottomrule
    \end{tabular}
    \caption{Results from the human 
    studies,
    showing proportional annotator preference for our system against the skyline.}
    \vspace{-0.1cm}
    \label{tab:listening}
\end{table}

\subsection{Human Reading Study}
While assessing lead sheet quality based on an audio rendering is useful as music is
experienced aurally, lead sheets are generally intended to be \emph{read} by musicians rather than learned by ear.
Furthermore, sheet music allows an experienced musician to 
quickly
analyze a piece 
based on learned visual patterns and compositional conventions
without having to frequently relisten.
Therefore to supplement the listening study, we also conducted a round of human evaluation in which annotators were asked to rank the lead sheets based on the sheet music alone.
In some ways this can be considered a closer simulation of the eventual downstream use case for this task.
For this \emph{reading} study we consulted 6 professional musicians, and presented them with 5 4--bar excerpts, also taken from the SOD validation set.
They were similarly asked to rank Lead-AE against skyline based on both accuracy and fluency.

On the right half of Table~\ref{tab:listening} we show the results from this evaluation.
We see that human experts once again preferred our system, and by an even larger margin on the criteria of accuracy.
An average of 80.0\% of annotators agreed with the majority on accuracy, and 56.7\% for fluency.
When asked to justify their answers, annotators often stated that the skyline omitted important melodic or harmonic information which Lead-AE was better able to capture.
The results of this reading study along with the listening study above corroborate our findings via the automatic metrics and indicate that Lead-AE is capable of generating significantly more accurate and functional lead sheets compared to the skyline baseline.

\section{Conclusion}
In this work we have presented Lead-AE, a novel architecture for modeling conditional lead sheet generation without relying on paired lead sheet--full score data.
On both automatic and human evaluations, Lead-AE consistently improves upon the determinstic baseline, and thus may inspire new directions for generating higher quality lead sheets and improve the downstream quality of larger arrangement systems.

\vfill\pagebreak

\bibliographystyle{IEEEbib}
\bibliography{strings}

\end{document}